\documentclass[prb,aps,floats,twocolumn,showpacs,groupedaddress,floatfix,amsmath]{revtex4-1}

\def\bc{\begin{center}}
\def\ec{\end{center}}
\def\beq{\begin{equation}}
\def\eeq{\end{equation}}
\def\bw{\begin{widetext}}
\def\ew{\end{widetext}}
\def\bea{\begin{eqnarray}}
\def\eea{\end{eqnarray}}
\def\non{\nonumber}

\def\dag{\dagger}

\def\ep{\epsilon}

\def\th{\theta}
\def\pa{\partial}

\renewcommand{\vec}[1]{\mbox{\boldmath$#1$}}
\usepackage{amssymb}
\usepackage{graphicx}
\usepackage{dcolumn}
\usepackage{bm}
\usepackage{subfigure}
\usepackage{array}
\usepackage{multirow}
\usepackage{booktabs}
\usepackage{appendix}
\usepackage{nicefrac}

\begin{document}

\title{Microscopic study of the $2/5$ fractional quantum Hall edge}
\author{G. J. Sreejith$^1$, Shivakumar Jolad$^{1,2}$,  Diptiman Sen$^3$, and 
Jainendra K. Jain$^{1}$}
\affiliation{$^1$Department of Physics, Pennsylvania State University, 
University Park, PA 16802}
\affiliation{$^2$Department of Physics, Virginia Tech, Blacksburg, Virginia 24061}
\affiliation{$^3$Center for High Energy Physics, Indian Institute of Science, 
Bangalore 560 012, India}

\date{\today}

\begin{abstract}
This paper reports on our study of the edge of the $\nicefrac{2}{5}$ fractional quantum Hall state, which is more complicated than the edge of the $\nicefrac{1}{3}$ state because of the presence of a continuum of quasi-degenerate edge sectors corresponding to different partitions of composite fermions in the lowest two $\Lambda$ levels. The addition of an electron at the edge is a non-perturbative process and it is not a priori obvious in what manner the added electron distributes itself over these sectors. We show, from a microscopic calculation, that when an electron is added at the edge of the ground state in the $[N_1,N_2]$ sector, where $N_1$ and $N_2$ are the numbers of composite fermions in the lowest two $\Lambda$ levels, the resulting state lies in either $[N_1+1,N_2]$ or $[N_1,N_2+1]$ sector; adding an electron at the edge is thus equivalent to adding a composite fermion at the edge. The coupling to other sectors of the form $[N_1+1+k,N_2-k]$, $k$ integer, is negligible in the asymptotically low-energy limit. This study also allows a detailed and substantial comparison with the two-boson model of the $\nicefrac{2}{5}$ edge. We compute the spectral weights and find that while the individual spectral weights are complicated and non-universal, their sum is consistent with an effective two-boson description of the $\nicefrac{2}{5}$ edge. 
\end{abstract}

\maketitle

\section{Introduction}

An understanding of the physics of the edge has been one of the long-standing challenges in the field of fractional quantum Hall (FQHE) systems [\onlinecite{Chang3}]. In ideal FQHE systems the excitations in the bulk are suppressed at low temperatures because of a gap, and low-energy excitations exist only at the edge. Furthermore, the electrons at the edge move only in one direction defined by the $\vec{E}\times\vec{B}$ drift. The edge thus behaves like a chiral one-dimensional system. Given the success of bosonization methods (see [\onlinecite{VonDelft}] for review) in dealing with one-dimensional electron liquids, it is attractive to attempt a description of the physics of the FQHE edge by reformulating the theory in terms of the bosonic density wave excitations following the usual method of bosonization. Such a theory requires $n$ species of bosons [\onlinecite{WenBlok}] for the FQHE states at $\nicefrac{n}{(2n+1)}$. Of its many predictions, the one that has been subjected to the most reliable experimental tests relates to the non-Ohmic behavior of the tunnel conductance for transport from an ordinary Fermi liquid into the FQHE edge [\onlinecite{Chang1,Grayson1,Chang2,Grayson2, Grayson3,Hilke}]. The bosonic approach predicts a power-law behavior $I\propto V^3$ for all FQHE states of the form $\nicefrac{n}{(2n+1)}$. Experiments do find a non-Fermi liquid behavior, but with an exponent of $\sim$ 2.8, 2.3 and 2.0 for $\nicefrac{1}{3}$, $\nicefrac{2}{5}$ and $\nicefrac{3}{7}$, respectively, and also do not find a plateau in the exponent as a function of the filling factor. A number of theoretical studies have explored the origin of the discrepancy [\onlinecite{zulicke96,Joglekar,OpenLiquids,DenserEdge,MomOccpDiscrp,MandalJain,ExactGreenFns,EdgeRecon,Instability,Yang03,Khve,LopezFradkin,LeeWen,Zulicke,contiVignale,DamicoVignale,JoladJain,JoladSenJain,NonFermionStatistics,PalaciosMacDonald}].

An independent approach for describing the FQHE edge uses the idea of composite fermions (CFs) [\onlinecite{JainCFpaper}],
without making any reference to bosonization. The microscopic foundations of the CF theory have been confirmed for the bulk physics (by comparison with experiment, or with exact results in the compact spherical geometry that contains no edges), and also for Hall droplets with an edge [\onlinecite{GunSang,Kamilla,Dev,Kawamura,JainCFBook}]. The FQHE state at $\nicefrac{n}{(2n+1)}$ is described as a state with $n$ filled $\Lambda$ levels, where 
the $\Lambda$ levels of composite fermions are analogous to the Landau levels of electrons, but reside within the lowest electronic Landau level. The edge excitations of this state then have a one-to-one correspondence 
with the edge excitations of the IQHE state with $n$ filled Landau levels. The natural questions that occur here are whether the CF theory and the bosonization approach are consistent, and what the precise correspondence is between the two pictures.

The most studied edge is that of the $\nicefrac{1}{3}$ state [\onlinecite{PalaciosMacDonald,JoladJain,JoladSenJain}]. This state consists of all the CFs occupying the lowest $\Lambda$ level, and its edge excitations mainly have all the CFs staying in the lowest $\Lambda$ level. Much less investigated, from a microscopic view point, are the edges of other fractions, although some work has been done in that direction [\onlinecite{MandalJain,ZueMacDonaldJohn,NonFermionStatistics}]. We focus in this paper on the simplest nontrivial edge, namely the edge of the $\nicefrac{2}{5}$ state, and seek to understand its physics from a microscopic starting point. The physics of the edges of other FQHE states of the form $\nicefrac{n}{(2n+1)}$ is expected to be similar.

A fundamental aspect in which the edge of the $\nicefrac{2}{5}$ system differs from the simpler $\nicefrac{1}{3}$ edge is the presence of several sectors in the edge spectrum at $\nicefrac{2}{5}$. Different sectors in the spectrum corresponds to different numbers of composite fermions in each of the two $\Lambda$ levels. Each sector has its own edge excitations, thus resulting in a fan-like diagram for edge excitations, as seen in Fig.~\ref{SpectraSectors} below. Several of these sectors are essentially degenerate in the thermodynamic limit for a realistic geometry and confinement. The same is of course true of the IQHE state at $\nu=2$ as well. As far as labeling and counting of edge modes is concerned, $\nu=\nicefrac{2}{5}$ is analogous to $\nu=2$.

However, a potential subtlety arises with extending the analogy between the FQHE and the IQHE to the process of adding an electron to the edge, which is relevant for the experiments quoted above. Suppose we begin with the ground state in the $[N_1,N_2]$ sector at $\nu=2$, which contains $N_1$ electrons in the lowest Landau level and $N_2$ in the second, and add an electron to this system. For a non-interacting system of electrons, we will end up with an excited state in either of the two sectors $[N_1+1,N_2]$ or $[N_1,N_2+1]$. Now consider $\nu=\nicefrac{2}{5}$, and add an {\em electron} at the edge of the ground state in the $[N_1,N_2]$ sector, in which $N_1$ composite fermions compactly occupy the lowest $\Lambda$ level and $N_2$ the second. The crucial point is to remember that we are adding an electron, not a composite fermion. As this electron gets converted into a composite fermion, which is a nonperturbative process, it is not clear, a priori, how it will spread over the states of all possible sectors of the form $[N_1+1+k,N_2-k]$, where $k$ is an arbitrary integer. One of the pleasing outcomes of our study is that, even in this case, the resulting state lies in either the $[N_1+1,N_2]$ or the $[N_1,N_2+1]$ sector. At low energies, adding an electron at the edge of a $\nu=\nicefrac{2}{5}$ system is thus equivalent to adding a composite fermion into one of the two $\Lambda$ levels.

An important simplifying assumption made in our work is that we neglect $\Lambda$ level mixing; the question of how $\Lambda$ level mixing modifies our conclusions is beyond the scope of the present work. There are reasons to believe that $\Lambda$ level mixing may be relevant. In a previous work the TL exponent of the edge liquid computed [\onlinecite{MandalJain}] by a direct evaluation of the equal-time Green's function was found to be in agreement with the prediction of the bosonic approach when $\Lambda$ level mixing was neglected, but was found to change when $\Lambda$ level mixing (caused by the residual interaction between composite fermions) was allowed. These studies indicate that the bosonized description is appropriate for noninteracting composite fermions, but inter composite fermion interactions produce corrections. It is not known what such corrections correspond to in the bosonic description. In this study we do {\it not} allow 
mixing with higher $\Lambda$ levels, and all our conclusions are subject to 
this assumption.

One of our aims in the present study is to establish a ``dictionary" between the operators and states in the effective two-boson description. A possible direct connection between them identifies the ``fermion" creation operator of the bosonic theory as the operator that actually creates a ``composite fermion" at the edge, with the bosonic degree of freedom representing particle-hole pairs of composite fermions at the edge. Such an identification can be verified by comparing the appropriately normalized matrix elements of the creation operators (which are called spectral weights, and which are precisely the elements that enter into the expression of the standard spectral function) in the two frameworks, following Palacios and MacDonald [\onlinecite{PalaciosMacDonald}]. Such tests have been performed [\onlinecite{PalaciosMacDonald}] for $\nu=\nicefrac{1}{3}$, and show that while the individual spectral weights are nonuniversal and do not necessarily conform to the bosonic model [\onlinecite{JoladJain}], their sum at a given momentum does [\onlinecite{JoladSenJain}]. The reason for considering the sum of the spectral weights at a given momentum is that the power-law exponent characterizing the tunnel conductance is determined in the asymptotic regime by the sum of the spectral weights for edges for which the dispersion is linear [\onlinecite{JoladSenJain}].

We calculate the sum of such spectral weights at various momenta (which correspond to angular momenta in our geometry) and compare our results to the E2BD description. The two ``bosons" represent bosonic particle-hole excitations of composite fermions at the edges of the two $\Lambda$ levels. The following facts support this conclusion:

\begin{enumerate}

\item The E2BD of the $\nicefrac{2}{5}$ edge contains two bosons. This is naturally expected within the CF theory where the $\nicefrac{2}{5}$ state has two filled $\Lambda$ levels, and thus two bosonic excitations associated with their edges.

\item The counting of edge excitations of CF $\Lambda$ levels matches with the counting from the E2BD.

\item Addition of an electron at the $\nicefrac{2}{5}$ edge is equivalent to the addition of a composite fermion at the edge, which can go into either of the two $\Lambda$ levels or some linear combination of them.

\item An explicit evaluation of the spectral weights (for transitions into the two relevant sectors) from the microscopic CF theory demonstrates that they satisfy the sum rule predicted by the E2BD.

\end{enumerate}

The plan of our paper is as follows. Section \ref{sec:CFDescription} contains a discussion of the CF description of the $\nicefrac{2}{5}$, and its ground states and excitations in different sectors. In Sec. \ref{sec:Bosonization}, we summarize the effective boson description of the edge of a $\nu=\nicefrac{2}{5}$ system. Section IV defines the spectral weights both in the E2BD and the microscopic theories, and also the value of the spectral weight sum predicted by the bosonic picture. Numerical results for the edge spectrum and spectral weights are presented in Sec. \ref{sec:num}. Section 
\ref{sec:results} presents the results and discussions. The paper is 
concluded in Sec. \ref{sec:conclusion}.

\section{Composite fermion theory of $\nu=\nicefrac{2}{5}$ edge}
\label{sec:CFDescription}

\begin{figure}
\includegraphics[scale=0.35]{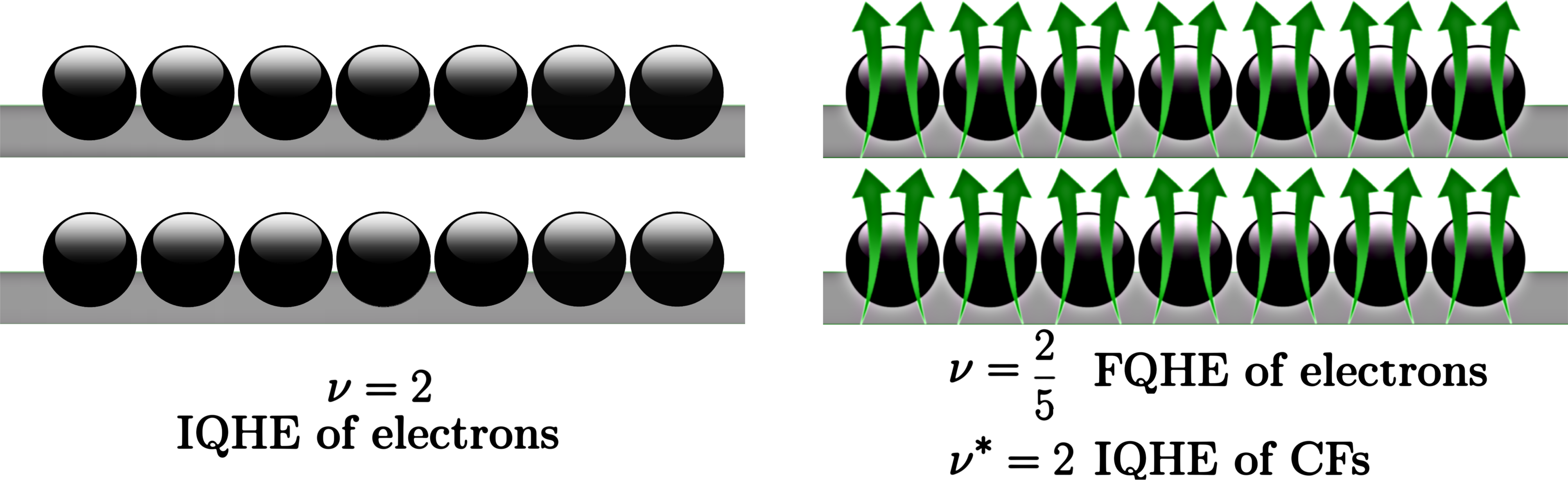}
\caption{(Color online) Schematic of $\nu=\nicefrac{2}{5}$ state in the composite fermion theory. Only 
the bulk structure is shown.} \label{TFSchematic} \end{figure}

FQHE results from the formation of quasiparticles called composite fermions (CFs) [\onlinecite{JainCFpaper}]. A composite fermion is a bound state of an electron and even number of quantized vortices. Strongly interacting electrons in a FQHE system are mapped to a system of weakly interacting CFs in an integer quantum Hall (IQH) state [\onlinecite{JainCFBook}]. A large class of the observed fractions $\nu=\nicefrac{n}{(2np\pm1)}$ ($n,p$ are integers), called Jain series, can be described in terms of an IQHE state $\nu^*=n$ of CFs with $2p$ vortices attached to them. The CFs in these states sense a lower magnetic field ($B^*$) than that of their electronic counterpart ($B$) given by, $B^*=B-2p\rho \phi_0$. Here $\rho$ is the density of electrons and $\phi_0=\nicefrac{hc}{e}$ is the magnetic flux quantum. The mapping between an electron FQHE state $\nu=\nicefrac{2}{5}$ and a CF $\nu^*=2$ state is illustrated in Fig.~\ref{TFSchematic}. A number of numerical studies have confirmed the validity of CF theory for the bulk physics of FQHE systems [\onlinecite{JainCFBook,KamillaJain,Heinonen}].

\begin{figure}
\includegraphics[scale=0.8]{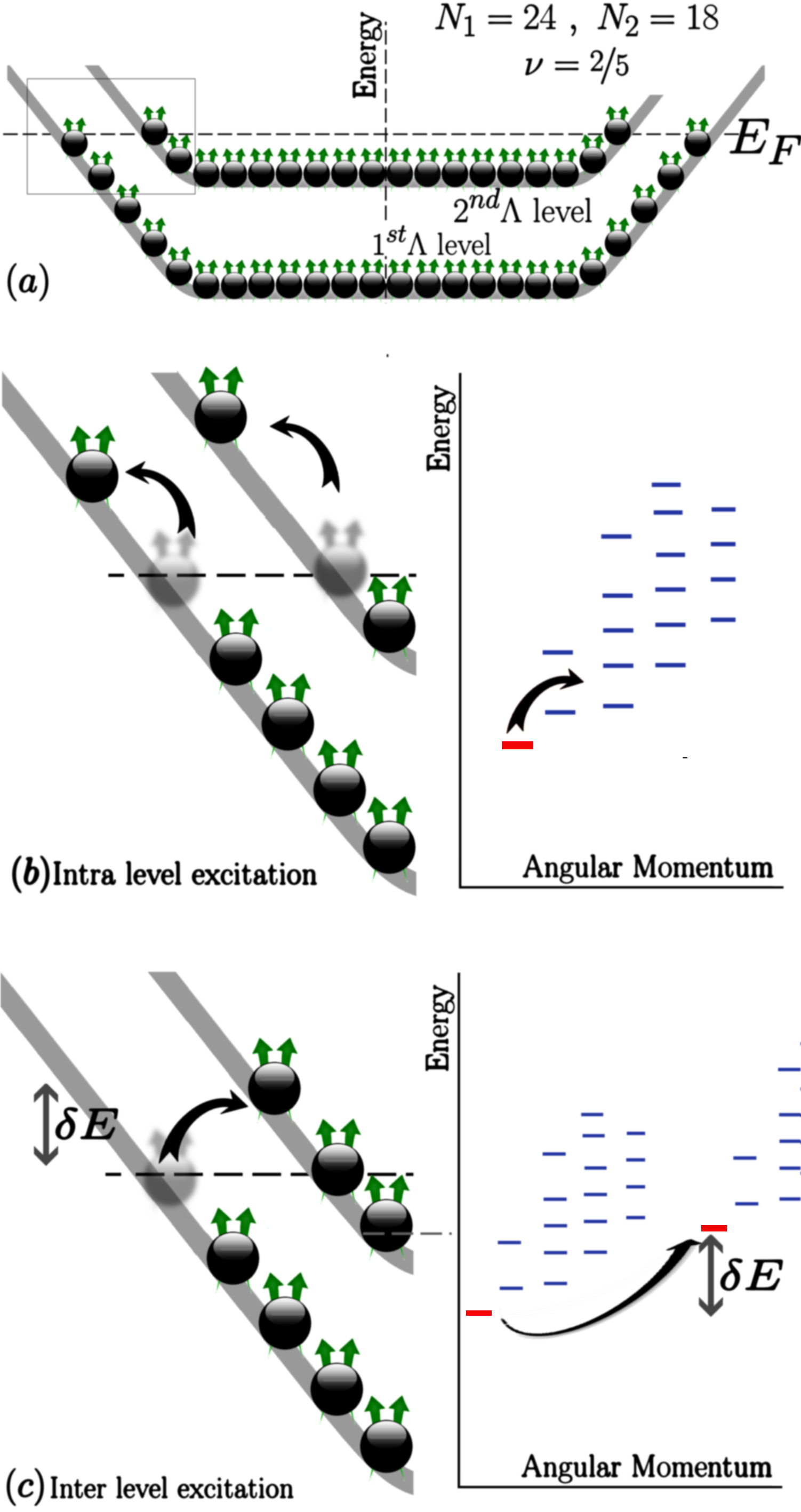}
\caption{(Color online) A schematic diagram of $\nu=\nicefrac{2}{5}$ edge. Panel (a) shows a compact ground state for the sector with $N_1$ and $N_2$ CFs in the lowest and second $\Lambda$ levels respectively. Excitations at the edge can be intra $\Lambda$ level or inter $\Lambda$ level. In an intra $\Lambda$ level excitation (panel b), the number of electrons in the individual $\Lambda$ levels do not change. In an inter $\Lambda$ level excitation (panel c), the electron makes a transition from one $\Lambda$ level to another, connecting the state to a neighboring sector as shown in the accompanying spectrum.}
\label{Schematic-edge} \end{figure}

In contrast to the case of $\nu=\nicefrac{1}{3}$ state which has all particles in the lowest $\Lambda$ level, the physics at the edge of the $\nicefrac{2}{5}$ system is complicated by the fact that it supports many ``sectors." A sector $[N_1,N_2]$ (with $N=N_1+N_2$) corresponds to states in which the first and second $\Lambda$ levels contain $N_1$ and $N_2$ CFs respectively (see Fig.~\ref{Schematic-edge}).
Within any given sector, the ``sector-ground-state'' is formed by composite fermions compactly occupying each $\Lambda$ level with $N_1$ and $N_2$ CFs respectively:
\beq
|\Psi_{[N_1,N_2]}^0\rangle= |[0,1,\cdots,N_1-1] [-1,0,\cdots,N_2-2]\rangle,
\eeq
where the numbers in the expression on the right hand side give the angular 
momentum quantum numbers of the CFs in the first and second $\Lambda$ levels.
The total CF angular momentum of this state is given by 
\beq M_0^* = \frac{N_1(N_1-1)}{2} + \frac{(N_2-2)(N_2-1)}{2}-1, 
\label{CFangmom} \eeq
and the corresponding electron angular momenta is $M_0=M_0^*+N(N-1)$. The lowest energy compact state occurring in an actual experiment is determined by the confinement potential, the interaction energy, and the total number of particles. 

The excited states within one sector consist of angular momentum excitations at the edge within each $\Lambda$ level. Such a state can be labeled by $\{{\bold n}_1,{\bold n}_2\}$ where ${\bold n}_i={(n_{i1},n_{i2}\dots)}$ describes the number of particles $n_{im}$ that are excited by an angular momentum $m$, and 
$i=1,2$ is the $\Lambda$ level index. For the excited states 
\beq M=M_0+\sum_m (n_{1m}+n_{2m})m.
\label{CFangmom-intra-excitation} \eeq
As can be seen from Fig.~\ref{Schematic-edge}, inter $\Lambda$ level 
excitations from the
ground state of one sector to another are accompanied by a large change of the
total angular momentum, whereas the intra $\Lambda$ level excitations can occur with very small angular momentum change.

\begin{figure}
\includegraphics[scale=0.45,viewport=0 0 600 350,clip]{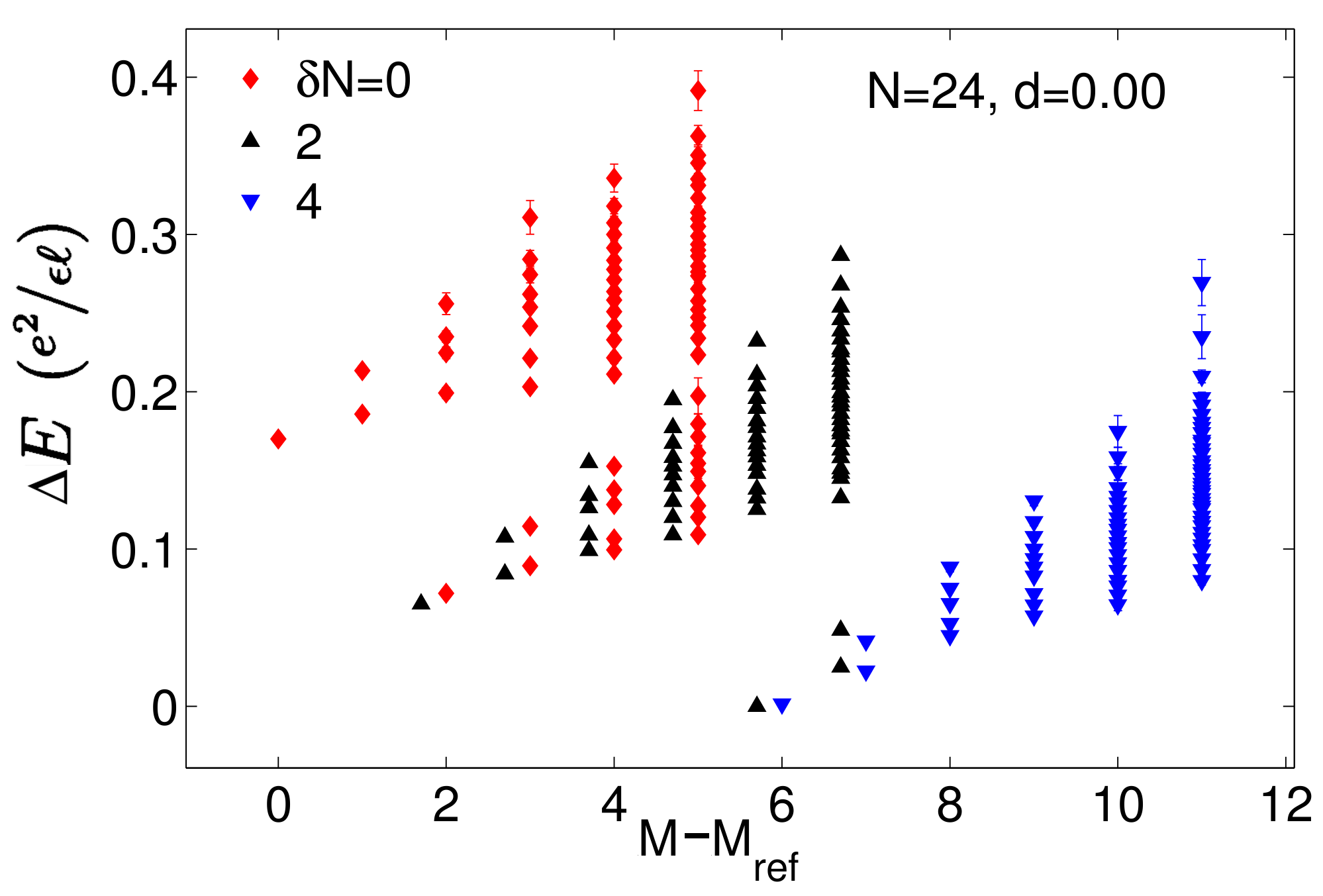}
\caption{(Color online) Energy spectrum of different sectors, $[N_1,N_2]=[(N+\delta N)/2, (N-\delta N)/2]$, of $\nicefrac{2}{5}$th state obtained through CF diagonalization. The spectrum is evaluated for the Coulomb interaction, in the presence of a neutralizing positively charged disk that also provides a confinement potential. The energy is quoted in units of $e^2/(\ep \ell)$, where $\ell$ is the magnetic length and $\ep$ is the background dielectric constant; the zero of energy is set at the ground state of the $\delta N=0$ sector. The following parameters are chosen: $N=24$, $\delta N=0,2,4$, setback distance $d=0.0$. The different symbols are slightly offset horizontally to avoid clutter. Angular momentum measured relative to the sector-ground-state of $[N/2,N/2]$}
\label{SpectraSectors}
\end{figure}

\begin{figure}
\includegraphics[scale=0.44]{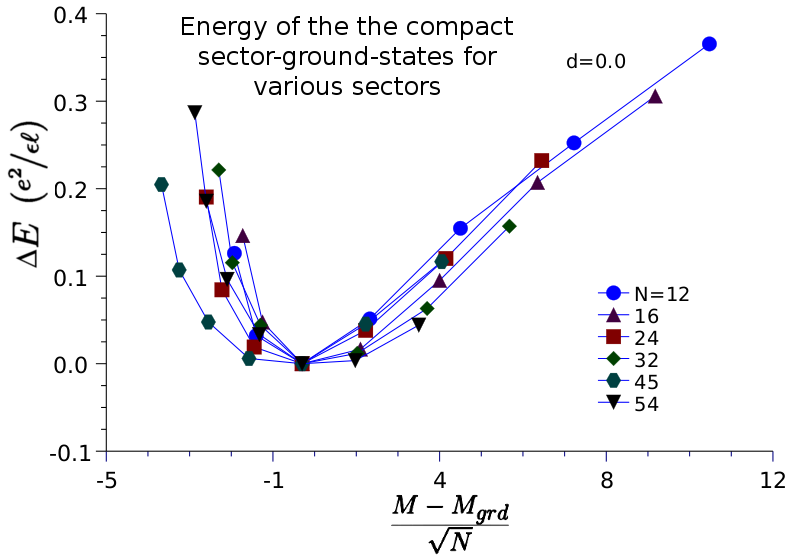}
\caption{(Color online) Energies of the compact ``ground" states in various sectors as a function of the total number of particles $N$. These occur at different total angular momenta in different sectors. The energies are evaluated for the Coulomb interaction, in the presence of a neutralizing positively charged disk that provides a confinement potential. The main message of this figure is to demonstrate that for large $N$, the curve flattens near the minimum, indicating that there are many almost degenerate ground states. $M_{\rm grd}$ is the angular momentum of the actual ground state among all the sectors.}
\label{SpectCompactStates}
\end{figure}

Within each sector, there are edge excitations of composite fermions that do not change the sector, as shown in Fig.~\ref{Schematic-edge} (b). This results in a number of ``fans" of edge excitations belonging to different sectors, as shown in Fig.~\ref{SpectraSectors}. (The spectra in this figure are obtained by the method of CF diagonalization, described below, in the presence of a uniform positively charged background that provides a confinement potential.) The {\em inter} $\Lambda$ level excitations correspond to transitions between different sectors. Such inter $\Lambda$ level excitations are suppressed in the bulk because of the gap, but such a gap does not exist at the edge, as seen explicitly in Fig.~\ref{SpectCompactStates}, which depicts the energies of the compact CF states as a function of $N$; it is clear that for large $N$ there is a continuum of almost degenerate sectors. As a result, excitations across different sectors exist at arbitrarily low energies and must be considered. 

\begin{figure}
\includegraphics[scale=0.46]{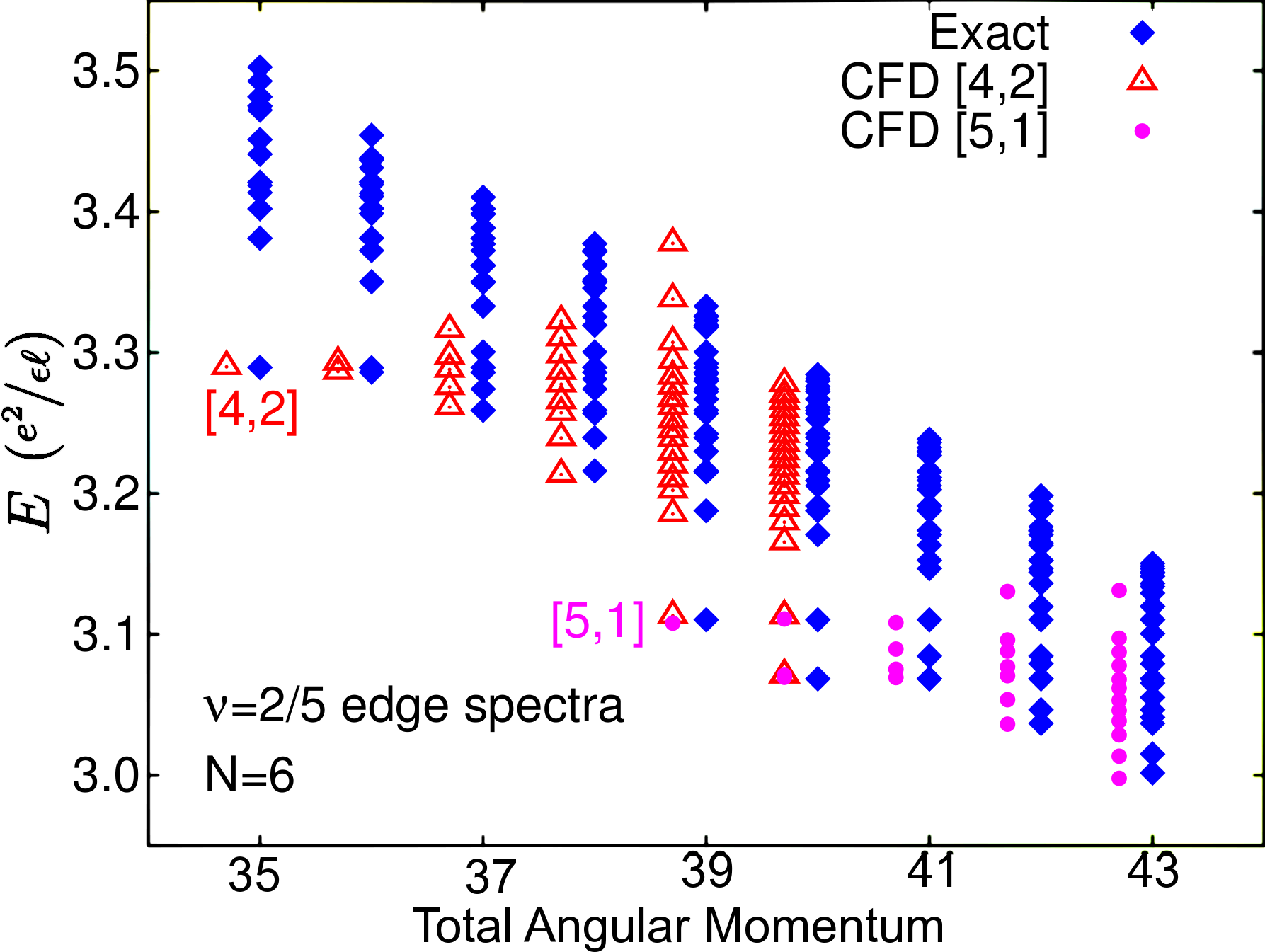}
\caption{(Color online) Comparison of CF diagonalization (CFD) and exact diagonalization (Exact) edge spectra for $N=6$ particles. The spectra are obtained in the absence of any neutralizing background. (The quality of the agreement persists even in the presence of the neutralizing background.) Energies are measured in units of $e^2/\ep l$. CFD energies are slightly offset from the exact energies along horizontal axis for clarity. The CFD spectra are obtained for two situations, with 4 and 2 composite fermions in the lowest two $\Lambda$ levels, and with 5 and 1 composite fermions in the lowest two $\Lambda$ levels; these spectra are shown with different symbols, as indicated on the plot.}
\label{CFDEDcompare}
\end{figure}

In order to assess the reliability of the subsequent results it is important to test how accurate the CF theory is for the edge excitations. In Fig.~\ref{CFDEDcompare}, we compare the Coulomb edge spectra of $\nu=\nicefrac{2}{5}$ for $N=6$ particles obtained through CF diagonalization. In obtaining the CF results, we do not include any $\Lambda$ level mixing (i.e. neglect CF transitions to higher $\Lambda$ levels), and we have also not included electron-background or background-background interactions. This figure also displays the exact spectra obtained from an exact diagonalization of the Coulomb interaction in the full lowest Landau level. A close agreement between the two spectra is evident at low energies.

Finally, we note that in Fig.~\ref{SpectraSectors} the spectrum of the edge excitations emanating from $[4,2]$ contain the edge excitations of $[5,1]$.  This, however, is relevant only for excitations with angular momenta (relative to the angular momentum of the ground state in a given sector) larger than the angular momentum difference between the ground states of the two sectors. Because the latter grows with $N$, such large angular momenta are not relevant to the edge physics in the thermodynamic limit. We always work with system sizes and angular momenta where such overlaps between sectors are not an issue.

\section{Effective 2-boson description of $\nicefrac{2}{5}$ edge}
\label{sec:Bosonization}

In this section, we will develop a two-boson theory for the edge of a 
quantum Hall system with filling fraction $\nu =\nicefrac{2}{5}$. The motivation for 
introducing {\it two} bosons comes from the CF description of such 
a system [\onlinecite{JainCFBook}]. If two flux quanta are attached to each
electron to form a CF, the electronic system with $\nu =\nicefrac{2}{5}$
effectively turns into a CF system with $\nu =2$, i.e., an
integer quantum Hall system with two filled $\Lambda$ levels, and therefore
two sets of chiral edge modes moving in the same direction. The bosonized
version of this system would therefore have two chiral bosons; these would
describe particle-hole excitations in the two edge channels, but they are not
capable of describing particle-hole excitations across the two edge channels.

Let us call the two bosonic fields $\phi_1$ and $\phi_2$ and assume that they 
are both right moving, going from $x=-\infty$ to $\infty$. We will not assume 
{\it a priori} that $\phi_1$ and $\phi_2$ separately describe the 
particle-hole excitations at the edges of the first and the second CF Landau 
levels; it is possible that both edge modes will involve some linear 
combinations of 
$\phi_1$ and $\phi_2$. The Lagrangian for the bosonic fields has the form
\bea L = \frac{1}{4\pi} \int_{-\infty}^\infty dx [ &-& \pa_t \phi_1 \pa_x \phi_1 - \pa_t \phi_2\pa_x \phi_2 \non \\
&-& \sum_{i,j=1}^2 \pa_x \phi_i V_{ij} \pa_x 
\phi_j ], \label{lag2}\eea
where $V_{ij}$ is a symmetric matrix, and we have absorbed the velocities $v_i$ in the definitions of the diagonal parameters $V_{ii}$. For repulsive density-density interactions, all the elements $V_{ij}$ will be positive. The momentum conjugate to $\phi_i$ is $-(1/4\pi) \pa_x \phi_i$. The charge operator of this theory takes the form [\onlinecite{Hansson}]
\beq Q = \frac{1}{4\pi} \int dx [ \frac{1}{\sqrt 3} \pa_x \phi_1 + 
\frac{1}{\sqrt 15} \pa_x \phi_2 ]. \eeq
The numerical factors in the above equation are justified by the observation
that $\nicefrac{1}{3} + \nicefrac{1}{15} = \nicefrac{2}{5}$, leading to the correct value of the Hall conductance.

Now we can see what form an electron annihilation operator must take. While it is not obvious from the effective boson approach, this operator will actually annihilate a composite fermion (i.e., an electron and two vortices). Therefore we will use that language, although it will be justified only later. Let us assume a form like
\beq \psi = \eta e^{-i (a_1 \phi_1 + a_2 \phi_2)},\eeq
where $\eta$ denotes a Klein factor which satisfies $\eta^\dag \eta = \eta 
\eta^\dag = 1$, and we have suppressed an overall normalization factor. For $\psi$ to anticommute with itself at different spatial 
points, we require $a_1^2 + a_2^2 =$ odd integer. (Note that the two-point 
correlation function of $\psi$ falls off as $\nicefrac{1}{|x-y|^{(a_1^2 + a_2^2)}}$). Next, 
in order for this to describe a CF with unit charge, we require
$[Q,\psi] = - \psi$, i.e., $a_1/\sqrt{3} + a_2/\sqrt{15} = 1$. At first sight, 
it seems that many different choices of $a_1, a_2$ are possible. However, if 
we want to have {\it two} operators with two pairs of values $a_1, a_2$ such 
that they mutually anticommute, and both satisfy $a_1^2 + a_2^2 =3$
(the latter condition ensures that the correlation function falls off
with the power 3 just as for a quantum Hall system with $\nu=\nicefrac{1}{3}$), 
then there are only two choices possible [\onlinecite{Hansson,HanssonHermans}]. Let us define the operators
\bea \psi_1 &=& \eta_1 e^{-i \sqrt{3} \phi_1}, \non \\
\psi_2 &=& \eta_2 e^{-i (2/\sqrt{3}) \phi_1 - i (\sqrt{5/3}) 
\phi_2 }, \label{psi12} \eea
where the Klein factors satisfy $\{ \eta_1 , \eta_2 \} = 0$. The effect of
$\eta_1$ and $\eta_2$ is to decrease the CF number by one in the first
and second Landau levels respectively. One can then show that $\psi_1$ 
and $\psi_2$ are both valid CF annihilation operators (i.e., each of them
carries unit 
charge and anticommutes with itself), and they also anticommute with each 
other due to the Klein factors. Further, each of them has a two-point 
correlation falling off as $\nicefrac{1}{|x-y|^3}$; the correlation between the two is 
zero since $\langle \eta_1^\dag \eta_2\rangle = 0$ in any eigenstate of the fermion 
number operator. 

Let us now assume that the electron annihilation operator is given by a 
linear superposition of the form
\beq \psi_e = c_1 \psi_1 + c_2 \psi_2, \label{psiexp} \eeq
where $c_1$, $c_2$ are some complex numbers.
We will also assume that the bosonic field on edge $i$ has the expansion
\bea \phi_{i+} (\th) &=& - \sum_{m > 0} \frac{1}{\sqrt m} b_{im}^\dag 
e^{im\th}, \non \\
\phi_{i-} (\th) &=& - \sum_{m > 0} \frac{1}{\sqrt m} b_{im} e^{-im\th},
\label{boson} \eea
where $m$ denotes the angular momentum of a bosonic mode; here we have 
assumed the edge to be the circumference of a circle and we have 
parametrized points on the edge by an angle $\theta$ going from 0 to $2\pi$.

\section{Spectral weights}

The theoretical quantity relevant to tunneling of an electron into the fractional quantum Hall edge is the spectral 
function of the edge. We concentrate below on the so-called spectral weights, which are the matrix elements that enter the expression of the spectral function 
(see Appendix A) [\onlinecite{PalaciosMacDonald}]. Furthermore, the density of states at a given energy is proportional to the 
sum of spectral weights for all states at that energy, which, for 
for bosons with linear dispersion, amounts to the sum over all states at the corresponding momentum.  The latter is 
easier to calculate theoretically (because the energy is a complicated function of various parameters), and therefore 
we will focus on the spectral weight sum over all states at a fixed (angular) momentum. 

\subsection{Bosons}

The spectral weights are defined as $|C_{\{{\bold n}_1,{\bold n}_2\}}|^2$, with
\beq 
C_{\{{\bold n}_1,{\bold n}_2\}}={\langle{\bold n}_1,{\bold n}_2|
\psi_e^\dag | 0 \rangle \over \langle 0|\psi_e^\dag|0\rangle}. \label{defineSW} 
\eeq
Here, $|0\rangle$ denotes the bosonic ground state and $|{\bold n}_1,{\bold n}_2\rangle=|\{ n_{11}, n_{12}, \cdots \};\{ n_{21}, n_{22}, 
\cdots \}\rangle$ an excited Fock state, where 
$n_{im}$ is the boson occupation number for the angular momentum $m$ state of 
the bosonic field $i$. The only role of the denominator is to cancel the (unknown) normalization 
factor in the definition of the electron field operator. 
The angular momentum (relative to the ground state) and energy of the 
this state are given by
\bea q &=& \sum_{m>0} (n_{1m} + n_{2m})m, \non \\
\Delta E &=& \sum_{m>0} (v_1 n_{1m} + v_2 n_{2m})m, \label{dme} \eea
where $v_i$ denotes the velocity of mode $i$; we are assuming a linear dispersion on each edge as is appropriate for a massless bosonic theory, but 
the velocities on the two modes can, in general, be unequal. 

Using Eqs.~(\ref{psi12}-\ref{boson}) in the definition of the spectral weight, we find the following expression for the spectral weight.
\beq C_{\{{\bold n}_1,{\bold n}_2\}}=\frac{{\langle 0|\prod_{i,m}
b_{im}^{n_{im}}\ \psi_e^\dagger|0\rangle}}{\sqrt{\langle 0|\psi_e
\psi_e^\dagger|0\rangle \langle 0|\prod_{i,m} b_{im}^{n_{im}}
b_{im}^{\dagger {n_{im}}}|0\rangle}} \eeq

This can be evaluated to get 
\bea
\vert C_{\{{\bold n}_1,{\bold n}_2\}}\vert^2 &=&|c_1|^2 \prod_j\frac{3^{n_{1j}}}{n_{1j}! j^{n_{1j}}} \prod_k 
\delta_{0,n_{2k}} \non \\&& +\;|c_2|^2 \prod_j \frac{(4/3)^{n_{1j}}}{n_{1j}!j^{n_{1j}}}\prod_k 
\frac{(5/3)^{n_{2k}}}{n_{2k}!k^{n_{2k}}}
\label{E2BDSumOfSW}
\eea
Table~\ref{swtable} lists the spectral weights for various states with 
$q=0$ to $3$.

\vskip .1 true cm
\begin{center}
\begin{table}[h]
{\renewcommand {\arraystretch}{1.7}
\begin{tabular}{|c|c|c|c|c|}
\hline
$q$ & $\left\{ n\right\} $ & spectral weight & partial sums & sum\\
\hline
0 						& $\{ 00,00\} $ 	& 	$|c_1|^2+|c_2|^2$ 				&	$1$ & $1$\\ 
\hline
\multirow{2}{*}{$1$} 	& $\{ 10,00\} $ 	& 	$3|c_1|^2+\frac{4}{3}|c_2|^2$	& \multirow{2}{*}{$3$} & \multirow{2}{*}{$3$}\\
						& $\{ 00,10\} $ 	& 	$\frac{5}{3}|c_2|^2$ &&\\
\hline

\multirow{5}{*}{$2$} 	& $\{ 20,00\} $ 	& 	$\frac{9}{2}|c_1|^2+\frac{8}{9}|c_2|^2$ & \multirow{3}{*}{ $\frac{9}{2}$ } & \multirow{5}{*}{$6$}\\

						& $\{ 00,20\} $ 	& 	$\frac{25}{18}|c_2|^2$ &&\\
						& $\{ 10,10\} $ 	& 	$\frac{20}{9}|c_2|^2$ &&\\
						\cline{4-4}
						& $\{ 01,00\} $ 	& 	$\frac{3}{2}|c_1|^2+\frac{2}{3}|c_2|^2$ & \multirow{2}{*}{$\frac{3}{2}$} &\\
						& $\{ 00,01\} $ 	& 	$\frac{5}{6}|c_2|^2$ &&\\
\hline

\multirow{10}{*}{$3$} 	& $\{ 30,00\} $ 	& 	$\frac{9}{2}|c_1|^2+\frac{32}{81}|c_2|^2$ & \multirow{4}{*}{$\frac{9}{2} $} & \multirow{10}{*}{$10$}\\
						& $\{ 00,30\} $ 	& 	$\frac{125}{162}|c_2|^2$ &&\\
						& $\{ 20,10\} $ 	& 	$\frac{40}{27}|c_2|^2$ &&\\
						& $\{ 10,20\} $ 	& 	$\frac{50}{27}|c_2|^2$ &&\\
						\cline{4-4}
						& $\{ 11,00\} $ 	& 	$\frac{9}{2}|c_1|^2+\frac{8}{9}|c_2|^2$ & \multirow{4}{*}{$\frac{9}{2} $} &\\
						& $\{ 00,11\} $ 	& 	$\frac{25}{18}|c_2|^2$ &&\\
						& $\{ 10,01\} $ 	& 	$\frac{10}{9}|c_2|^2$ &&\\
						& $\{ 01,10\} $ 	& 	$\frac{10}{9}|c_2|^2$ &&\\
						\cline{4-4}
						& $\{ 001,000\} $ 	& 	$|c_1|^2+\frac{4}{9}|c_2|^2$ & \multirow{2}{*}{$1$} &\\
						& $\{ 000,001\} $ 	& 	$\frac{5}{9}|c_2|^2$ &&\\
\hline

\end{tabular}
\caption{The last column shows the sum of spectral weights for excitations with angular momenta $q =0,1,2$ and $3$, while setting $|c_1|^2+|c_2|^2=1$. The various states at each $q$ are shown in the second column, in the notation explained in the text. The third column shows the individual spectral weights, and the fourth column gives the partial sums, where each sum comes from states in which a given number of bosons are excited.}
\label{swtable}

}
\end{table}
\end{center}
\vskip .2 true cm

Certain sum rules can be gleaned from the above table.
If we add up all the spectral weights for a given value of $q$, we
obtain $[(q + 1)(q + 2)/2]~(|c_1|^2 + |c_2|^2)$; this is just 
as in the case of $\nu = \nicefrac{1}{3}$ as given in Palacios and MacDonald 
[\onlinecite{PalaciosMacDonald}]. This makes sense since both our 
CF operators $\psi_1$ and $\psi_2$ are analogous to the CF operator 
for $\nu = \nicefrac{1}{3}$ in every way, i.e., they have the same scaling 
dimension ($=3$) and the same kind of expansion in terms of bosons. We also 
observe some finer partial sum rules within each value of $q$. For 
instance, within $q = 2$, the first three states add up to $(9/2) 
(|c_1|^2 + |c_2|^2)$, while the last two states add up to $(3/2) (|c_1|^2 + 
|c_2|^2)$. These are exactly what we find in Table II of 
Ref.~[\onlinecite{PalaciosMacDonald}], where we see the numbers 9/2 and 3/2 for
the states $\{2000\}$ and $\{0100\}$ for $\nu = \nicefrac{1}{3}$. A similar
statement holds for the states with $q = 3$. In general, there is a partial sum rule for all the states with same total angular momentum and with same total number of bosons.

\subsection{Composite fermions}
\label{subsec:SW}

An {\em electron} added at the edge of the ground state $\Psi_{[N_1,N_2]}^0$ of sector $[N_1,N_2]$ can distribute itself into the available CF states of the form $\Psi^{\{{\bold n}_1,{\bold n}_2 \}}_{[N_1',N+1-N'_1]}$ in different sectors of the $N+1$ system. Within each sector, the counting of edge states agrees with the E2BD, and it is natural to ask which sectors are relevant, and whether the sum rule is satisfied in these sectors.
Motivated by previous studies on $\nu=\nicefrac{1}{3}$ [\onlinecite{PalaciosMacDonald, JoladJain, JoladSenJain}], we define the spectral weights for such a process as $\left|C_{\{ {\bold n}_1,{\bold n}_2 \}} \right|^2$, where
\beq C_{\{{\bold n}_1,{\bold n}_2\}}= \frac{ \langle 
\Psi_{[N_1',N_2']}^{\{{\bold n}_1,{\bold n}_2\}} | a_{m_0+q}^\dagger|
\Psi_{[N_1,N_2]}^0\rangle}{\langle \Psi_{[N_1',N_2']}^{0} | a_{m_0}^\dagger|
\Psi_{[N_1,N_2]}^0\rangle}. \label{defineSW} \eeq
Here $a_m^\dagger$ adds an electron in the lowest {\it Landau} level at an angular momentum $m$. The quantity $m_0$ is the angular momentum difference between the sector-ground-states of $[N_1,N_2]$ and $[N_1',N_2']$. We note that because of the difference in the ground state angular momenta in the different sectors, the spectral weights at angular momentum $q$  in E2BD correspond to the addition of an electron at angular momentum $m_0+q$ in the electronic language. However, we continue to call this the ``spectral weight at angular momentum $q$" for ease of comparison with the E2BD picture.
The states $a_m^\dagger|\Psi_{[N_1,N_2]}^0\rangle$, $|\Psi_{[N_1',N_2']}^{\{{\bold n}_1,{\bold n}_2\}} \rangle$ and $|\Psi_{[N_1,N_2]}^{\{{\bold n}_1,{\bold n}_2\}} \rangle$ are normalized. Although not explicitly shown in the notation, the spectral weights depend on the sectors $[N_1,N_2]$ and $[N_1',N_2']$ of the initial and final states.

For a choice of the orthonormal basis $\Psi_{[N_1',N_2']}^{\{{\bold n}_1,{\bold n}_2\}}$, the sum of the spectral weights at a given angular momentum $m$ within a given sector is
\bea
S_q =\sum |C_{\{{\bold n}_1,{\bold n}_2\}}|^2 = \sum \left|\frac{ \langle \Psi_{[N_1',N_2']}^{\{{\bold n}_1,{\bold n}_2\}} | a_{m_0+q} ^\dagger|\Psi_{[N_1,N_2]}^0\rangle}{\langle \Psi_{[N_1',N_2']}^{0} | a_{m_0}^\dagger|\Psi_{[N_1,N_2]}^0\rangle} \right|^2 = \non \\
\frac{\langle \Psi_{[N_1,N_2]}^0 |a_{m_0+q} \left[\sum|\Psi_{[N_1',N_2']}^{\{{\bold n}_1,{\bold n}_2\}}\rangle \langle \Psi_{[N_1',N_2']}^{\{{\bold n}_1,{\bold n}_2\}} |\right] a_{m_0+q}^\dagger |\Psi_{[N_1,N_2]}^0\rangle}{\left|\langle \Psi_{[N_1',N_2']}^{0} | a_{m_0}^\dagger|\Psi_{[N_1,N_2]}^0\rangle\right|^2}. \non 
\eea
The sum inside the square bracket is the projection $\mathcal{P}^q_{[N_1',N_2']}$ into the space of all states within the given sector and of angular momentum $m_0+q$. So the sum of spectral weights can be conveniently written as
\bea S_q=\frac{\left| \mathcal{P}^q_{[N_1',N_2']}a^\dagger_{m_0+q} |
\Psi_{[N_1,N_2]}^0\rangle \right|^2}{\left|\langle \Psi_{[N_1',N_2']}^{0} | 
a_{m_0}^\dagger|\Psi_{[N_1,N_2]}^0\rangle\right|^2}. \label{projection} \eea
The sum of the spectral weights is therefore related to the part of the new state $a^\dagger_{m_0+q}|\Psi_{[N_1,N_2]}^0\rangle$ that lies in the sector of the CF states under consideration.

The physics that we wish to verify is that the above annihilation operator in fact describes the annihilation of a CF at the edge of the quantum Hall system, and that two bosons span the space of CF particle-hole excitations at the edge of the two $\Lambda$ levels. A microscopic verification of this physics is achieved by comparing the sum of the spectral weights predicted from the bosonization approach with that of the numerically calculated sum. Note that the bosonic 
operators $b_{1m}$ and $b_{2m}$ do not exactly correspond to excitations at the two edges; namely, subsets of excitations of the two bosons and excitations at 
the two edges may be related to each other by unitary transformations which 
are unknown and which
may be different for different subsets. Hence the partial sum rules cannot be 
explicitly verified in the CF theory.

In what follows, we neglect $\Lambda$ level mixing, and only consider edge excitations within a given sector. Furthermore, we do not include any confinement potential or any electron-electron interaction. These are not relevant for the total spectral sum rule which is invariant under an unitary rotation of the basis; with our neglect of $\Lambda$ level mixing, the only role of the confinement potential or the Coulomb interaction is to produce a different basis. 

\section{Numerical methods}
\label{sec:num}

\subsection*{Model for Energy calculations}

The system is modeled as a quantum Hall droplet in a disk geometry, described previously in Ref.[\onlinecite{JoladSenJain,JoladJain}]. For completeness, we give a brief outline here. Electrons are confined to a disk of radius $\sqrt{2N/\nu}$ magnetic lengths by a uniformly distributed neutralizing background positive charge located at a setback distance of $d=0$ from the disc. The Hamiltonian for this system is
\bea H_I &\equiv& V_{\rm ee}+V_{\rm eb}+V_{\rm bb} \non \\
&=& \sum_{j<k} \frac{e^2}{\ep |\vec{r}_j-\vec{r}_k|} - \rho_0\sum_{j} \int_{\Omega_N} d^2r \frac{e^2}{\ep \sqrt{|\vec{r}_j-\vec{r}|^2}} \non \\
&& + \rho_0^2 \int_{\Omega_N} \int_{\Omega_N} d^2rd^2r'\frac{e^2}{\ep 
|\vec{r}'-\vec{r}|}, \label{Hami} \eea
where the terms on the right hand side represent the 
electron-electron, electron-background, and background-background energies, 
respectively. Here $\vec{r}_j$ is the position of the $j^{th}$ electron, $\rho_0= \nu/2\pi l^2$ 
is the positive charge density spread over the disc, and 
$\ep$ is the dielectric constant of the background semiconductor material. The kinetic energy term in not considered explicitly as only the lowest Landau level states are occupied at high magnetic field. 

In order to obtain the exact spectrum, the above Hamiltonian must be diagonalized in the Hilbert space of all the $N$-electron states in the lowest Landau level. Due to exponentially growing dimension of this space, it becomes impractical to compute the spectra for systems containing more than $\sim$ 10 electrons. However, since we are interested only in the low energy features of the spectra, a very accurate description is obtained by diagonalizing the above Hamiltonian in the basis of CF states, described in the following paragraph, the dimension of which is much smaller than that of the electron basis, thus enabling us to study much larger systems. Access to larger systems is crucial for obtaining the thermodynamic limits shown below. It is expected that the universal properties of the edge will not depend sensitively on the precise form of the wave function, so our CF states ought to be adequate. (We note that even the exact electron states will depend on the shape of the confining potential, finite thickness corrections to the interaction, LL mixing, etc.)

For the fraction $\nu=\nicefrac{n}{(2np+1)}$, the CF theory maps interacting electrons at total angular momentum $M$ to non-interacting composite fermions at $M^*= M-pN(N-1)$ [\onlinecite{Dev,Kawamura}] by attaching $2p$ vortices to each 
electron. The ansatz wave functions $\Psi^M$ for interacting electrons with angular momentum $M$ are expressed in terms of the known wave functions $\Phi^{M^*}$ of non-interacting electrons at filling fraction $n$ at total angular momentum $M^*$ as follows:
 \beq \Psi^M = {\cal P}_{\rm LLL} \prod_{j<k}(z_j-z_k)^{2p} 
 \Phi^{M^*}_\alpha \label{PsiAlpha}. \eeq 
where ${\cal P}_{\rm LLL}$ denotes projection into the lowest Landau level. In general, there are many different ways of partitioning the total angular momentum $M^*$ to the $N$ non-interacting fermions, thus producing several states at the same angular momentum (labeled by $\alpha$ in the preceding equation). These states span the low energy basis in which the Hamiltonian in Eq.\ref{Hami} is diagonalized. Fig.\ref{CFDEDcompare} shows a comparison of the spectrum obtained by diagonalizing the Coulomb interaction in the full Hilbert space as well as in the CF basis.

\subsection*{Numerical evaluation of sum of spectral weights}

\begin{figure}[h]
\includegraphics[scale=0.42]{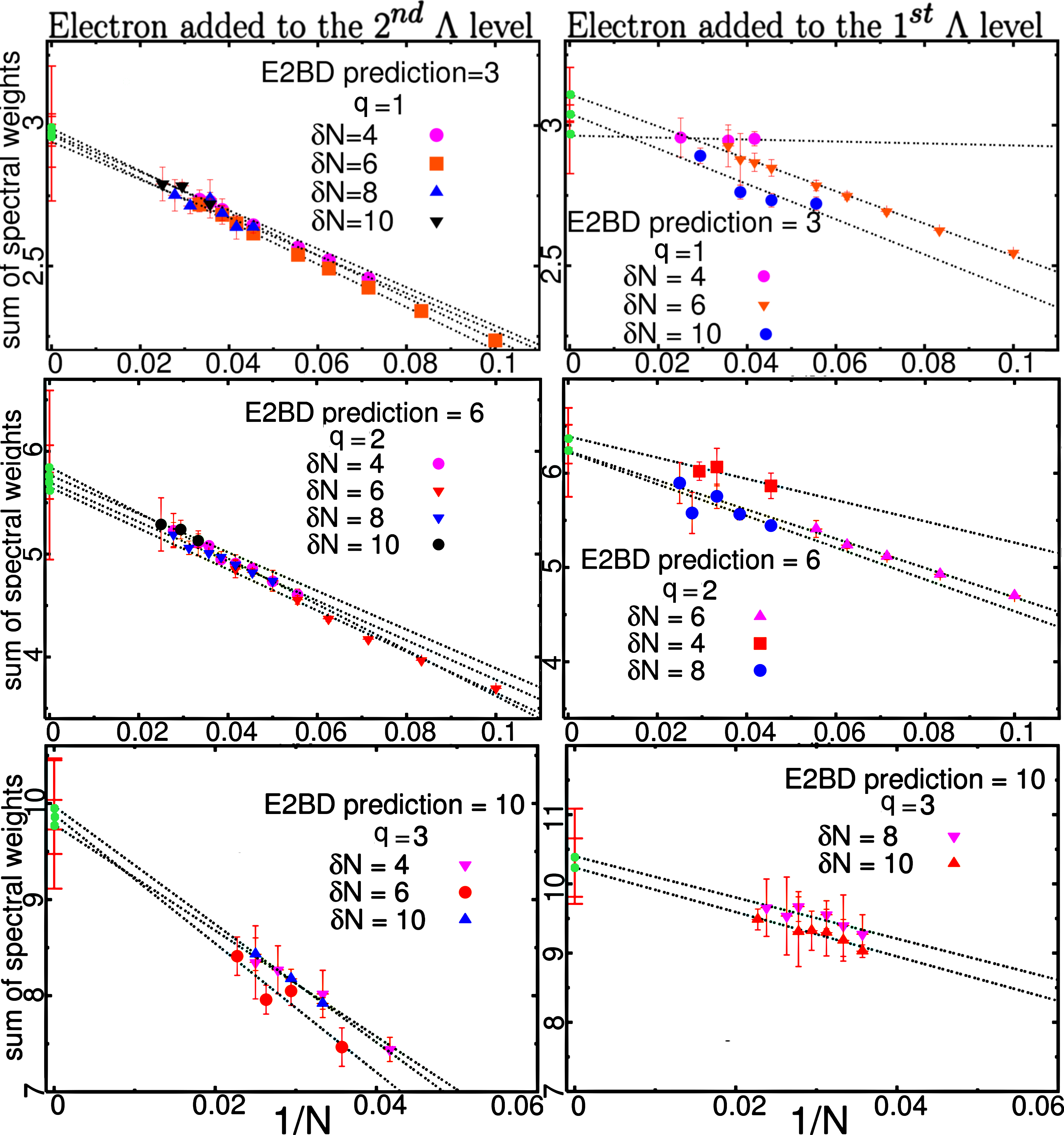}
\caption{(Color online) Spectral weight sums as a function of $N$ for $q=1,2$ and $3$. The columns on the right are for a composite fermion added to the lowest $\Lambda$ level, whereas those on the left are for a composite fermion added to the second $\Lambda$ level. The values predicted by the E2BD are shown in the figures. The thermodynamic limits of the sums are consistent with these values in all cases. Here $\delta N=N_1-N_2$.}
 \label{sw}
\end{figure}

The individual spectral weights $|C_{\{{\bold n}_1,{\bold n}_2\}}|^2$ depend on the states $\Psi_{[N_1',N_2']}^{\{{\bold n}_1;{\bold n}_2\}}$ that are chosen. However the sum of the spectral weights (Eq.~(\ref{projection})) is independent of this choice of basis. The projection operator can therefore be expanded using any convenient basis for the space of a given angular momentum. For the numerical calculations we choose a basis $\{\psi_{[N_1,N_2]}^j\}$ where the CFs occupy fixed orbitals. Here $j$ represents a composite index representing the angular momentum orbitals occupied by the CFs in the two $\Lambda$ levels. These states are not orthogonal to each other, and therefore the projection operator expanded using these states has to be written as 
\beq \mathcal{P}^m_{[N_1',N_2']}=\sum_{i,j}\left|\psi_{[N_1,N_2]}^i
\right\rangle [\mathcal{O}^{-1}]_{ij} \left\langle\psi_{[N_1,N_2]}^j\right|, 
\eeq
where the overlap matrix $\mathcal{O}$ is defined as
\beq \mathcal{O}_{ij}=\left\langle\Psi_{[N_1',N_2']}^i|\Psi_{[N_1',N_2']}^j
\right\rangle. \eeq
Plugging this into Eq. (\ref{projection}) gives 
\beq S={\bold C}^\dagger\mathcal{O}^{-1}{\bold C}, \eeq
where ${\bold C}$ is the column vector of ``unsquared'' spectral weights 
of these basis states
\beq C_i = \frac{ \langle \Psi_{[N_1',N_2']}^i| a_{m_0+q}^\dagger|\Psi_{[N_1,N_2]}^0
\rangle}{\langle \Psi_{[N_1',N_2']}^{0} | a_{m_0}^\dagger|\Psi_{[N_1,N_2]}^0
\rangle}. \label{sw-in-special-basis} \eeq
The quantities $C_i$ as well as $\mathcal{O}$ were evaluated numerically using Metropolis-Hastings Monte Carlo[\onlinecite{ChibMetropolis,Hastings,Metropolis}] integration algorithms. Several different values of $[N_1,N_2]$, with $N_2<N_1$ were chosen for this calculation. Systems with the different $\delta N=N_1-N_2$ appeared to follow slightly different paths to the thermodynamic limit when the sum was plotted against $1/N$. So extrapolations to the thermodynamic limit were done for systems with specific values of $\delta N$ as shown in Fig. \ref{sw}.

\begin{figure}[h]
\includegraphics[scale=0.83]{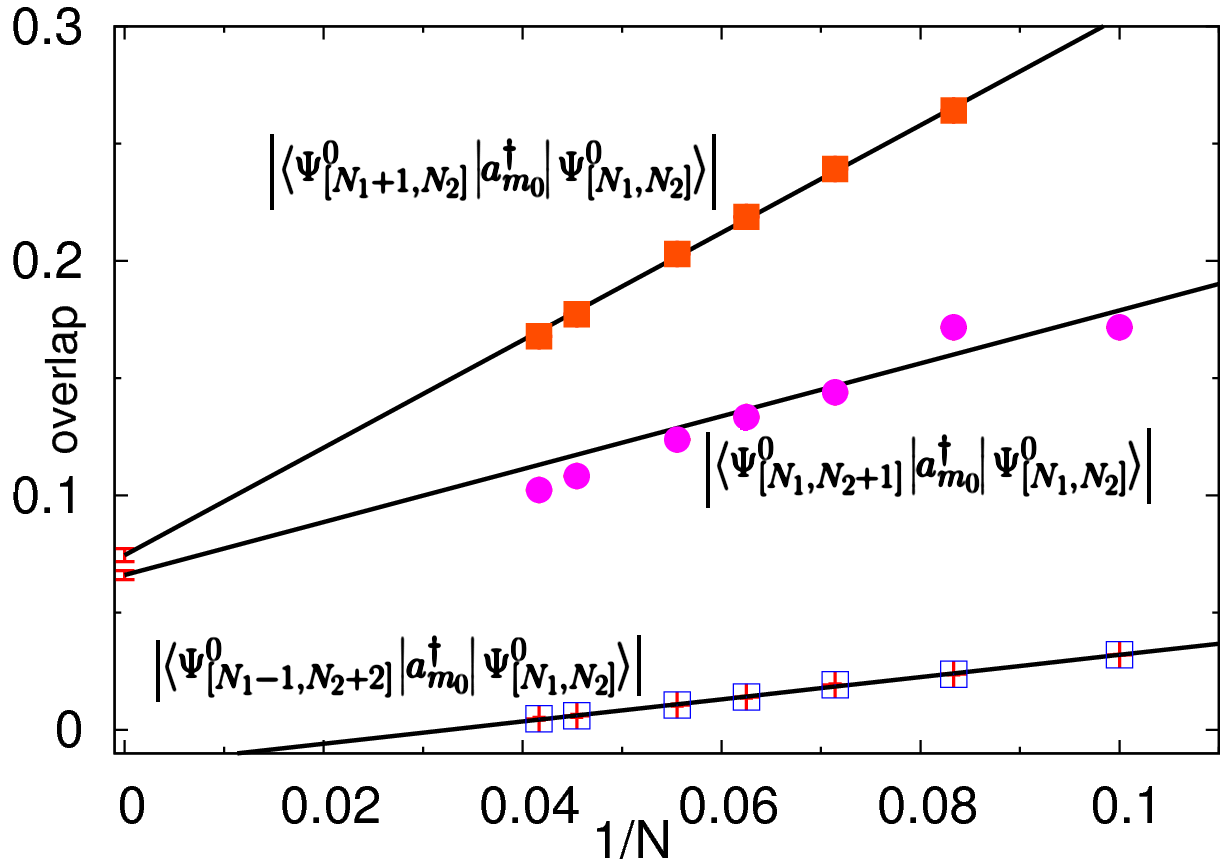}
\caption{(Color online) Overlap of the state obtained by adding a single electron to the ground state in the $[N_1,N_2]$ sector, namely $a_{m_0}^\dagger |\Psi^0_{[N_1,N_2]}\rangle$,  with various sector-ground-states.  Here $N_1-N_2=6$, and the states $a_{m_0}^\dagger |\Psi^0_{[N_1,N_2]}\rangle$ and $|\Psi^0_{[N_1+j,N_2-j+1]}\rangle$ are taken to be normalized. The coupling with the sector $[N_1-1,N_2+2]$ grows vanishingly small as the size of the system increases. As shown in the text, the coupling with the sector $[N_1+2,N_2-1]$ is identically zero by symmetry. We expect therefore that the coupling vanishes for all sectors other than $[N_1+1,N_2]$ and $[N_1,N_2+1]$, indicating that only these two sectors dominate tunneling of an electron into the edge of the $\nicefrac{2}{5}$ state.}
 \label{coupling}
\end{figure}

\section{Results and Discussion}
\label{sec:results}

Ideally, we would like to test the two-edge nature of the spectrum through the sum rule for $A(m_1,m_2)$ given in Eq. \ref{sumrule2}, which depends separately on the angular momenta $m_1$ and $m_2$ of the excitations in the two edge channels. However, we have found that it is not possible, perhaps because of finite size effects, to identify in our calculated spectra the eigenstates arising from the different edge channels. As a result, we concentrate below on the total sum rule for $A(m)=(m+2)(m+1)/3$ given by Eq. \ref{sumrule}.

Column $1$ of Fig.~\ref{sw} shows the sum of the spectral weights at constant total angular momenta assuming that the newly added electron goes entirely as a CF into the $2^{\mathrm{nd}}$ $\Lambda$ level, corresponding to the transition from sector $[N_1,N_2] \rightarrow [N_1,N_2+1]$. 
The individual spectral weights are all non-zero and do not match the numbers given in Table \ref{swtable}. It is expected that the individual spectral weights are not the same as the prediction since the bosonic operators $b_{im}$ do not directly correspond to the angular momentum excitations in the $\Lambda$ levels
(though they must be related to each other by unitary transformations). The sum of the spectral weights, however, matches the sum predicted by the E2BD, in 
the thermodynamic limit.

Column $2$ of Fig.~\ref{sw} shows the sum of the spectral weights for the case where the newly added electrons enters as a composite fermion in the lowest $\Lambda$ level producing a final state in the sector $[N_1+1,N_2]$. In this case, the numerical evaluation of the spectral sum rule can be simplified by noting that only one of the spectral weights is non-zero for the CF basis (as shown analytically in Appendix \ref{sw_in_lll}). The sum of the spectral weights is again in agreement with the E2BD predictions. The fact that the sum of the spectral weights at each angular momentum matches with the same from the bosonization approach shows that there is a one-to-one correspondence between the vector space of states with constant total angular momentum in the two approaches.

In addition to these two situations, one can consider the case where the added electron appears as a CF in one of the other sectors, such as $[N_1,N_2]\rightarrow [N_1-1,N_2+2]$.  Fig.~\ref{coupling} shows the overlap matrix element for several sector-ground-states. The overlap with sector-ground-state of $[N_1-1,N_2+2]$ rapidly approaches 0 as $N$ increases. Using simple angular momentum accounting, similar to the one in Appendix \ref{sw_in_lll}, one can show that the overlap of the new state into the sector $[N_1+2,N_2-1]$ is identically zero. These results indicate that the coupling to sectors other than $[N_1+1,N_2]$ and $[N_1,N_2+1]$ is vanishingly small; the other sectors can therefore be neglected insofar as the process of tunneling of an electron is concerned.

\section{Conclusions}
\label{sec:conclusion}

We have undertaken an investigation of the $\nicefrac{2}{5}$ edge within the framework of the microscopic CF theory. Our motivations are twofold. The $\nicefrac{2}{5}$ edge has interesting additional structure, due to the presence of multiple quasi-degenerate edge sectors, that is not found at the \nicefrac{1}{3} edge, and tunnel conductivity experiments exhibit a larger discrepancy from the predictions of the effective bosonic theory. 

We have found several interesting results. First of all, even though there is a continuum of quasi-degenerate fans of edge excitations, belonging to different $[N_1,N_2]$ sectors, the tunneling is dominated by two sectors, which are the sectors in which a composite fermion is added at the edge of the lowest or the second $\Lambda$ level. The addition of an electron to the edge of a FQHE system is thus equivalent to the addition of a composite fermion -- a result that is pleasing but far from obvious. Second, we find that the values of the spectral weight sum at constant total angular momentum is, in the thermodynamic limit, consistent with the values predicted by the effective two-boson theory. These sum rules govern the exponent relevant for the tunnel conductance at low biases; thus our results provide a nontrivial microscopic confirmation of the predictions of the two-boson theory (within our approximations). Strictly speaking, our analysis holds for a situation in which the two edge channels have the same velocity, so the sum over spectral weights at a fix momentum is identical to the sum over spectral weights at a fixed energy, but arguments can be given (see Appendix A) that the same exponent is obtained even when the edge channels have different velocities. A numerical verification of sum of spectral weights for constant momenta in the {\em individual} modes is yet to be accomplished. Finally, we confirm the operators constructed in Refs.~[\onlinecite{HanssonHermans}] and [\onlinecite{Hansson}] for adding composite fermions to various $\Lambda$ levels.

We conclude by speculating on the origin of the discrepancy between the E2BD predictions and the experimental results. One possibility is that of edge reconstruction [\onlinecite{EdgeRecon}] which renders the exponents non-universal. However, intrinsic sources for the discrepancy have not been ruled out. As indicated in Ref.~[\onlinecite{MandalJain}], $\Lambda$ level mixing, which is always present (because although the inter-CF interaction is small compared to the CF cyclotron energy, the two have the same energy scale) but neglected in the present work, can possibly be relevant and can produce corrections to the edge exponents. Further investigations will be required to sort out these effects.

\section{Acknowledgments}
We acknowledge Paul Lammert, Chuntai Shi, and Vikas Argod for insightful discussions and support with numerical codes and cluster computing. The work at Penn State was supported by the DOE under Grant No. DE-SC0005042.  S.J. acknowledges the partial support from National Science Foundation under Grants No. NSF-DMR-0705152 and 1005417. D.S. thanks DST, India for financial support under SR/S2/JCB-44/2010. The authors acknowledge Research Computing and Cyberinfrastructure, a unit of Information Technology Services at The Pennsylvania State University, for providing high-performance computing resources and services used for the computations in this work. 

\appendix

\section{Green's Functions for $\nu=\nicefrac{2}{5}$ }
\label{App:GF}

Here we derive the Green's function based on bosonic fields described in Sec. \ref{sec:Bosonization},
\bea
G(x,t) &=& \langle 0|T\{\psi(x,t)\psi^{\dagger}(0,0)\}\vert 0\rangle \non \\
&\propto& \sum_{ \{{\bold n}_1,{\bold n}_2\}} e^{i{\displaystyle 
\ep_{{\bold n}}t}} e^{-ik_{\bold n}x } \vert C_{\{{\bold n}_1,{\bold n}_2\}}\vert^2
\label{basicGF}
\eea
Here we have assumed $t>0$, and the momentum is taken as the total angular momentum $k_{\bold n}=m_1+m_2\equiv m$, where  
$m_{1}=\sum_{l}ln_{1l}$ and $m_{2}=\sum_{l'}l'n_{2l'}$ are the
angular momenta of electrons in the first and second $\Lambda$ levels respectively. 

If we assume a linear dispersion for both edge channels with the same velocity, 
then the energy of the state $\vert {\bold n}_1,{\bold n}_2 \rangle$ is given by $\epsilon=vm$, which allows us to write an expression for the spectral function $A(m)$, 
defined by 
\beq 
G(x,t)=\sum_{m=0}^{\infty} A(m)e^{im(vt-x)},\label{GF:E=vm}
\eeq
as
\beq
A(m)=\sum_{ \{{\bold n}_1,{\bold n}_2\}} \vert C_{\{{\bold n}_1,{\bold n}_2\}} \vert^2\;\delta\left(m-{\textstyle \sum_l}(n_{1l}+n_{2l})l\right)\non
\eeq
In other words, the spectral function at a given momentum (or energy) is the sum over the spectral weights of all states at that momentum. 
E2BD predicts this sum to be 
\beq
A(m)={(m+2)(m+1)\over 2}.
\label{sumrule}
\eeq
This is directly related to the edge Luttinger liquid exponent. Inserting this in Eq.\ref{GF:E=vm}, the Green's function is seen to be
\beq
	G(x,t)=\sum_{m=0}^{\infty} \frac{(m+2)(m+1)}{2} e^{im(vt- x)}\approx\frac{1}{\vert x-vt\vert ^3}\non
\eeq

For situations where the two modes in the E2BD model have different velocities, as in general expected for a realistic situation, 
the energy of a state is given by $\epsilon=v_1m_1+v_2m_2$. The Green's function Eq.\ref{basicGF} can be written as 
\beq G(x,t)=\sum_{m_1=0}^{\infty} \sum_{m_2=0}^{\infty} A(m_1,m_2)e^{i(m_1+m_2)(vt- x)},\eeq
where $A(m_1,m_2)$ is the sum of spectral weights of states in which angular momenta in the two modes are $m_1$ and $m_2$ respectively.
\bea
&&A(m_1,m_2)=\non \\&&\sum_{ \{{\bold n}_1,{\bold n}_2\}} \vert C_{\{{\bold n}_1,{\bold n}_2\}} \vert^2\;\delta\left(m_1-{\textstyle \sum_l}n_{1l}l\right)\delta\left(m_2-{\textstyle \sum_l}n_{2l}l\right)\non
\eea
The E2BD prediction for spectral weights (eq.\ref{E2BDSumOfSW}) can be used to evaluate this sum. The primed summations below correspond to the sum over ${\bold n}_i$ such that $\sum_l ln_{il}=m_i$. 
\bea
A(m_1,m_2)&=&\vert c_1\vert^2\sum_{{\bold n}_1}'{\textstyle \prod_l}\frac{3^{n_{1l}}}{n_{1l}!j^{n_{1l}}} \sum_{{\bold n}_2}'{\textstyle \prod_l} \delta(n_{2l}-0)\non\\
&+&\vert c_1\vert^2\sum_{{\bold n}_1}'{\textstyle \prod_l}\frac{(\nicefrac{4}{3})^{n_{1l}}}{n_{1l}!j^{n_{1l}}} \sum_{{\bold n}_2}'{\textstyle \prod_l}\frac{(\nicefrac{5}{3})^{n_{2l}}}{n_{2l}!j^{n_{2l}}}\non\\
A(m_1,m_2)&=&\vert c_1\vert^2 \binom{m_1+2}{m_1} \delta(m_2-0) \non\\
&+&\vert c_1\vert^2 \binom{m_1+1/3}{m_1} \binom{m_2+2/3}{m_2}
\label{sumrule2}
\eea
where we have used the identity
\beq
	\sum_{{\bold n}}'{\textstyle \prod_l}\frac{\alpha^{n_{l}}}{n_{l}!j^{n_{l}}}=\binom{m+\alpha-1}{m},
\eeq
Using eq.\ref{sumrule2}, the leading terms of the Green's function can be evaluated to be
\beq
	G(x,t)\approx \frac{\vert c_1 \vert^2} {\vert  x-v_1t \vert ^3}+\frac{\vert c_2 \vert^2}{\vert  x-v_1t \vert ^{\nicefrac{4}{3}} \vert  x-v_2t \vert ^{\nicefrac{5}{3}}}\label{GF2:leading}
\eeq
While the corresponding spectral function at a fixed energy is more complicated, one can argue that in the low energy limit, which corresponds to the $t\to\infty$ limit, the above Green's function scales as $\nicefrac{1}{\vert t\vert^3}$, thus again producing the exponent of 3.

\section{Spectral weights for particle added to lowest $\Lambda$ level}
\label{sw_in_lll}
When an electron is added to the lowest $\Lambda$ level edge, the spectral weight at angular momentum $q$,
\beq
	C_i=\frac{ \langle\Psi^i _{[N_1+1,N_2]}|a^\dagger _{m_0+q}| \Psi^0 _{[N_1,N_2]}\rangle}{ \langle\Psi^0_{[N_1+1,N_2]}|a^\dagger _{m_0}| \Psi^0 _{[N_1,N_2]}\rangle}\non ,
\eeq
where $m_0=3N_1+N_2$ is the difference in angular momentum between the sector ground states at $[N_1,N_2]$ and $[N_1+1,N_2]$, 
can be shown to vanish for most cases. In fact, it is nonzero only for 
\beq
|\Psi^i _{[N_1+1,N_2]}\rangle=	\left |[0,1,2\dots , N_1-1,N_1+q] [-1,0,1\dots , N_2-2]\right\rangle 
	\label{non-zero-sw}
\eeq
The result follows because the state $a^\dagger _{m_0+q}| \Psi^0 _{[N_1,N_2]}\rangle$ has the single particle orbital 
at angular momentum $(m_0+q)$ occupied with probability {\em one}. It is an easy exercise to check that the basis state 
shown in Eq.\ref{non-zero-sw} is the only CF state that has a nonzero occupation of that angular momentum orbital; all 
other CF basis states have that orbital unoccupied, thus producing a zero overlap.

\end{document}